\documentstyle[12pt,psfig,epsf]{article}

\begin{document}

\parskip=0.3cm
\begin{titlepage}

\hfill \vbox{\hbox{UNICAL-TH 06/4}
\hbox{UAB-FT-529}\hbox{June 2002}}

\vskip 0.2cm

\centerline{\bf EXPLICIT MODEL REALIZING PARTON-HADRON DUALITY$~^\star$}

\vskip 0.3cm

\centerline{R.~Fiore$^{a\dagger}$, 
A.~Flachi$^{b\diamond}$,
L.L.~Jenkovszky$^{c\ddagger}$,
A.I.~Lengyel$^{d\S}$, V.K.~Magas$^{c,e\ast}$}

\vskip 0.1cm

\centerline{$^{a}$ \sl  Dipartimento di Fisica, Universit\`a della Calabria 
\& INFN-Cosenza,}
\centerline{\sl I-87036 Arcavacata di Rende, Cosenza, Italy} 

\centerline{$^b$ \sl IFAE, Universidad Aut\`onoma de Barcelona,}
\centerline{\sl 08193 Bellaterra, Barcelona, Spain}

\centerline{$^{c}$ \sl  Bogolyubov Inst. for Theor. Phys., Ac. of Sciences 
of Ukraine}
\centerline{\sl UA-03143 Kiev, Ukraine}

\centerline{$^{d}$ \sl Institute of Electron Physics,}
\centerline{\sl Universitetska 21, UA-88000 Uzhgorod, Ukraine}

\centerline{$^e$ \sl Center for Physics of Fundamental 
Interactions (CFIF),}
\centerline{\sl Physics Department, Instituto Superior Tecnico,} 
\centerline{\sl Av. Rovisco Pais, 1049-001 Lisbon, Portugal}

\vskip 0.1cm

\begin{abstract}
An explicit model realizing parton-hadron duality and fitting the
data is suggested. Complex nonlinear Regge trajectories are important
ingredients of the model. The inclusion of $\Delta$ and 
$N^*$ trajectories should account for all resonances in the
direct channel. The exotic trajectory is responsible for the
smooth background.

\end{abstract}

\vskip 0.1cm

\hrule

\vskip 0.1cm

\noindent
$^{\star}${\it Work supported in part by the 
Ministero dell'Istruzione, dell'Universit\`a e della Ricerca (MIUR) and by 
the INTAS}

$
\begin{array}{ll}
^{\dagger}\mbox{{\it e-mail address:}} &
   \mbox{FIORE@CS.INFN.IT} \\
^{\ddagger}\mbox{{\it e-mail address:}} &
\mbox{JENK@GLUK.ORG} \\
^{\diamond}\mbox{{\it e-mail address:}} &
\mbox{FLACHI@IFAE.ES} \\
^{\S}\mbox{{\it e-mail address:}} &
   \mbox{SASHA@LEN.UZHGOROD.UA} \\
^{\ast}\mbox{{\it e-mail address:}} &
   \mbox{VLADIMIR@CFIF.IST.UTL.PT}
\end{array}
$

\end{titlepage}
\eject
\newpage

\textheight 210mm \topmargin 2mm \baselineskip=24pt


\newcommand{\dlt}{\bigtriangleup}
\newcommand{\beq}{\begin{equation}}
\newcommand{\eeq}[1]{\label{#1} \end{equation}}
\newcommand{\insertplot}[1]{\centerline{\psfig{figure={#1},width=15.0cm}}}
\newcommand{\insertplotsshort}[1]{\centerline{\psfig{figure={#1},height=5.0cm}}}
\newcommand{\insertplotll}[1]{\centerline{\psfig{figure={#1},height=11.0cm}}}
\newcommand{\insertplotlll}[1]{\centerline{\psfig{figure={#1},height=22.0cm}}}

\section {Introduction} \label{s1}
 The photoabsorption cross section in the resonance region has
 been studied in a large number of papers \cite{Kobberling, Stein, 
Brasse, Bodek, Whitlow}
 (for a comprehensive review see Ref.~\cite{BSYP}).
There are nearly 20 resonances in the $\gamma^* p$ system in the
region between the pion-nucleon threshold and below $2$ GeV, but
only a few of them can be identified more or less unambiguously.
One reason is that they overlap and compete with changing $Q^2$
and the other is the uncertainty due to the background.
Therefore, instead of identifying each resonance, one considers
three maxima above the elastic scattering peak,
corresponding to some ``effective'' resonance contributions.
Most of the data come from SLAC and have been compiled by Stoler
\cite{Stoler}. The first maximum is due to the isolated
$\Delta(1232)$ resonance. The second resonance region is dominated
by two strong negative-parity states, the $D_{13}(1520)$ and the
$S_{11}(1535)$. At low $Q^2 \ \ (<1$ GeV$^2$) the $D_{13}(1520)$
dominates, whereas at high $Q^2 \ \ (\sim 3$ GeV$^2$) there is some
evidence that the $S_{11}(1535)$ becomes dominant. 
The relative strength of the other states is
not well determined, especially at increasing $Q^2$. For example,
in the fits for $Q^2>4$ GeV$^2$ it is usually assumed that only
$S_{11}(1535)$ contributes to the second resonance, eventually
corrected for $Q^2\leq 4$ GeV$^2$ to subtract the contribution
from the $D_{13}(1520).$ In the third
resonance region, the strongest excitation at low $Q^2$ is the
$F_{15}(1680)$ state. Although it is assumed that
this remains one state, it is possible that several states
contribute strongly at high $Q^2.$ All the fits include a state at
$W=1440$ MeV, corresponding to the possible location of the
$P_{11}(1440),$ but these fits do not exhibit any positive
contribution from this state. Ultimately,
in most of the phenomenological fits only three peaks are
included, namely those at $W\sim 1232,\ \ 1535$ and $1680$ MeV.

The standard approach to the phenomenological analysis of the data
is that developed in Ref.~\cite{Kobberling} and still widely used.
Ignoring the (relatively small) longitudinal component of the
total cross section, one writes
 the contribution of the three prominent
resonances, 
$\sigma_T^R=\sigma_T^{\Delta}+\sigma_T^2+\sigma_T^3$, with a
relativistic Breit-Wigner formula for the $\Delta$ resonance and
non-relativistic formulae for the the second and the third
resonance regions. \footnote{We remind that this is a phenomenological
description since the Breit-Wigner formulae are applied to a peak
created by a superposition of resonances rather than to a single
resonance.}

The above resonance contribution is appended by a smooth
background, parametrized as 
\beq
\sigma_T^{bg}=\sum_{n=1}^3 C_n(Q^2)(W-W_{th})^{n-1/2},
\eeq{m1}
where the square-root term $(W-W_{th})^{1/2}$ takes into account 
the behavior of non-resonant pion production at the threshold
$W_{th}=m_p+m_\pi$ ($m_p$ is a proton mass). The coefficients $C_n(Q^2)$
are polynomials of $Q^2$: $C_n(Q^2)=\sum_{j=0}^4 C_{nj}Q^{2j}$.
A detailed analysis of the SLAC data along these lines can be
found e.g. in \cite{Bodek, Dissertation}.

Systematic theoretical studies of the subject, including the
helicity structure of the amplitude, the threshold- and the
QCD-motivated asymptotic behavior of the form factors can be
found in the papers by Carlson and co-authors \cite{CM, CPoor}
(see also Ref.~\cite{SS}). Spin structure functions (SF's) were studied 
in Ref.~\cite{spinCM}.

Recent results  from  JLab (CEBAF) \cite{Nicu}  renewed the
interest in the subject \cite{Armstrong, Ent, Simula} and they
call for a more detailed phenomenological analysis of the data and
a better understanding of the underlying dynamics.

In the present paper we develop further the arguments presented in
Refs. \cite{JMP,JMP-1,JMSz} and earlier works cited therein. The basic
idea in our approach is to use the off-mass-shell continuation of
the dual amplitude with nonlinear complex Regge trajectories.
These trajectories play a crucial role in the dynamics of the
strong interactions. Actually, the trajectories can be considered
as the basic dynamical variables, replacing the usual Mandelstam
variables $s,t$ and $u.$ In particular, their form determines
completely the spectrum of resonances (see e.g. Ref.~\cite{FJMPP}). The
parameters of the trajectories can be fitted independently of the
masses and widths of the known resonances, therefore, in
principle, they reflect more adequately the position of 
the peaks in $ep$ scattering, formed
by the interplay of different resonances. In concentrating on this
aspect of the dynamics, we leave more freedom to the choice of the
$Q^2$-dependent form factors. We start with a simplified model,
disregarding the helicity structure of the amplitudes. We ignore
the relatively small (and poorly known) contribution from the
cross sections involving longitudinally polarized photons,
$\sigma_L$ .
In doing so, we anticipate the connection \cite{JMP} with the
small-$x$ (high-energy) domain, where these simplifications are
commonly accepted.

Apart from the above-mentioned details, there is an important
problem in the $Q^2$-dependence, namely the asymptotic behavior of
the form factors, usually related to the quark counting rules.
While the validity of these rules for elastic form factors leaves a
little doubt, they may get modified for the transition form factors.
Moreover, in our approach a new type of form factors, generalizing
the concept of inelastic form factors, appears. These
new form factors correspond to
the transition of the nucleon to a baryon trajectory, with a
sequence of nucleon resonances on it. It follows from dual models
(see Section 3 and Ref.~\cite{JMP}) that the powers of these form
factors increase with the spin of the excited state.

We adopt the two-component picture of strong interactions
\cite{FH}, according to which direct-channel resonances are dual
to cross-channel Regge exchanges and the smooth background in the
$s-$channel is dual to the Pomeron exchange in the $t-$channel. As
explained in Ref.~\cite{JMP}, the background corresponds in a dual
model to a pole term with an exotic trajectory that does not
produce any resonance.

To summarize the philosophy of the present paper (for a related
approach see Ref.~\cite{Domokos}), we believe that the mechanism
of virtual photoproduction on nucleons is not much different from
that in purely hadronic processes. Since at intermediate energies
the entire nondiffractive part of the hadronic amplitude is
saturated by direct-channel resonances, by duality they determine
completely the dynamics of the reaction. By expressing the
inelastic electron scattering cross section through the imaginary
part of the virtual Compton scattering amplitude, it is reasonable
to assume that a resonance model will work for virtual Compton
scattering equally well as for pion-nucleon scattering, i.e. both
are saturated by the contribution of many overlapping resonances.
The crucial question is the spectrum of these resonances,
connected with the form of nonlinear, complex Regge trajectories.

The paper is organized as follows: 
the kinematics of $eN$ scattering and the relevant notation is described 
in Section~\ref{s2}.
The properties of Regge trajectories, which constitute a basic 
ingredient of our model, are discussed in Section~\ref{s3}. 
The model for the 
nucleon structure function is
presented in Section~\ref{s4} and the fitting procedure and
comparison with data is discussed in Section~\ref{s5}. 
Sections~\ref{s7} and
\ref{s8} are devoted to analytical and numerical tests of parton-hadron
duality. Finally, we draw our conclusions in Section~\ref{s9}. 

\section {Notation, kinematics and observables} \label{s2}

\begin{figure}[htb]
        \insertplotsshort{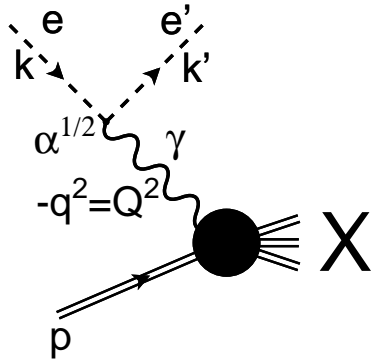}
\caption{Kinematic of deep inelastic scattering.} 
\label{d1}
\end{figure}

We study inclusive electron-nucleon scattering shown in Fig. \ref{d1}.
The scattered electron, entering into the process with 
energy $E$, emerges with energy $E'$ at an angle
$\theta$ with respect to the initial direction. 
The four-momentum transferred by the virtual photon
from the electron to the nucleon is $q=k-k'=-Q^2$, where $k$ and
$k'$ are the four-momenta of incoming and outgoing electrons,
respectively. The energy transferred between the electron and the
nucleon (or the energy lost by the electron) is
\beq
\nu={pq\over m}=E-E',
\eeq{m30}
where $p$ is the four-momentum of the target nucleon of mass $m$.
The invariant squared mass of the recoiling system (hadronic final
state) is
\beq
W^2=(p+q)^2=m^2+2m\nu-Q^2.
\eeq{m29}

 An experimentally measurable quantity in the the reaction
$e(k)+N(p)\rightarrow e'(k')+X$ is  the differential cross
section (see e.g. Ref.~\cite{Close})
\beq
{d^2\sigma^{\gamma^* p}\over{d\Omega dE'}}=\Gamma(\sigma_T+\epsilon\sigma_L),
\eeq{m28}
 where
\beq
\Gamma={\alpha K\over{2 \pi^2 Q^2 }}{E'\over
E}{1\over{1-\epsilon}},
\eeq{m27}
with $K=(W^2-m^2)/(2m)$ and
\beq
\epsilon=\left(1+2{\nu^2+Q^2\over{Q^2}}\tan^2{\theta\over
2}\right)^{-1}
\eeq{m26}
or
\beq
{d^2\sigma^{\gamma^* p}\over{d\Omega dE'}}=\Gamma \sigma_T(1+\epsilon R).
\eeq{m25}
The ratio $R=\sigma_L/\sigma_T$ relates the transverse component
of the cross section, $\sigma_T,$ to the longitudinal component,
$\sigma_L.$ The structure function $F_2$ can be expressed
in terms of the differential cross section and $R$ as
\beq
F_2={d^2\sigma^{\gamma^* p}\over{d\Omega dE'}}{1+R\over{1+\epsilon R}}
\left[\frac{K\nu}{4\pi^2\alpha}{1\over \Gamma}\frac{1}{1+{\nu^2 \over Q^2}}\right]\ ,
\eeq{m24}
where the kinematics is given by the term in square brackets.

Phenomenological parameterizations for the  quantity $R$ exist in
the literature \cite{Whitlow}, but in this paper, as we are ignoring 
the spin structure, we shall put $R=0$.

The central object of the present study is the nucleon SF, 
uniquely related to the photoproduction cross section by

\beq
F_2(x,Q^2)={Q^2(1-x)\over{4\pi \alpha (1+{4m^2 x^2\over {Q^2}})}}
\sigma_t^{\gamma^*p}(s,Q^2)\ ,
\eeq{m23}
where total cross section, $\sigma_t^{\gamma^* p}$, 
is the imaginary part of the forward Compton scattering
amplitude, $A(s, Q^2)$,  
\beq
\sigma_t^{\gamma^* p}(s)={\cal I}m\ A(s, Q^2)~.
\eeq{m22}
 The center of mass energy of the $\gamma^* p$ system,
 the negative squared photon virtuality $Q^2$ and the Bjorken
 variable $x$ are related by
 \beq
s=W^2=Q^2{(1-x)\over x}+m^2,
\eeq{m21}
Instead of  $W^2$, we use the Mandelstam variable $s$, typical of hadronic reactions.

\section {Nonlinear complex Regge trajectories.\\ Direct-channel resonances and background}
\label{s3}
\begin{figure}[htb]
        \insertplot{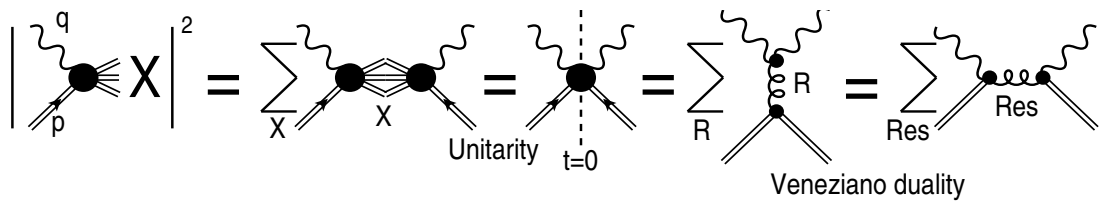}
\caption{According to the Veneziano (or resonance-Reggeon) 
duality a proper sum 
of either $t$-channel or $s$-channel resonance exchanges accounts for
the whole amplitude. } 
\label{d2}
\end{figure}

The SF is related via Eq.~(\ref{m23}) to the total cross
section, or the imaginary part of the forward Compton scattering
amplitude, Eq.~(\ref{m22}). 
The latter, on the other hand, is related by unitarity
to the sum of all possible intermediate states, as shown in 
Fig. \ref{d2} (see e.g.
Ref.~\cite{Collins}). One should distinguish between cross sections
summed over a limited number of nuclear excitations
\cite{Kobberling, Stoler}, $\sigma_{\gamma N\rightarrow R_i}$ and
the total cross section of virtual forward Compton scattering,
related to the SF, including by unitarity all
possible intermediate states allowed by energy and quantum number
conservation.

The correct way to fully account for all possible intermediate
states in the resonance region is in terms of the $s$-channel
Regge trajectories, which automatically include the huge
number of resonances as recurrences, appearing on the
trajectories \cite{JMSz}. 
This is an economic way to take into account the whole
sequence of nucleon excitations, lying on a single trajectory.
This is more than a mere technical simplification: Regge
trajectories are basic building blocks in dual models.
The kinematical variables enter through Regge trajectories,
which thus play the role of dynamical variables. Now, since
the behavior of the Regge trajectories is assumed to be known in
the whole range of their variation, the resulting behavior of the
SF can be thus extrapolated towards small $x$,
well beyond the resonance region.

The dynamics of the resonance formation is intimately related to
the form of the Regge trajectories. Their linear rise is a widely
accepted approximation. It is based on the observed spectrum of
resonance masses and was supported by narrow resonance dual
models and their mechanical analogues, such as the harmonic
oscillator or relativistic strings.

In fact, linear trajectories contradict both the theory and the
data: analyticity requires the presence of threshold singularities
in the trajectories, while their asymptotic behavior is also
constrained by an upper bound on their real part. The finite
widths of resonances also require the presence of a nonvanishing
imaginary part in the trajectories.

\begin{figure}[htb]
        \insertplotll{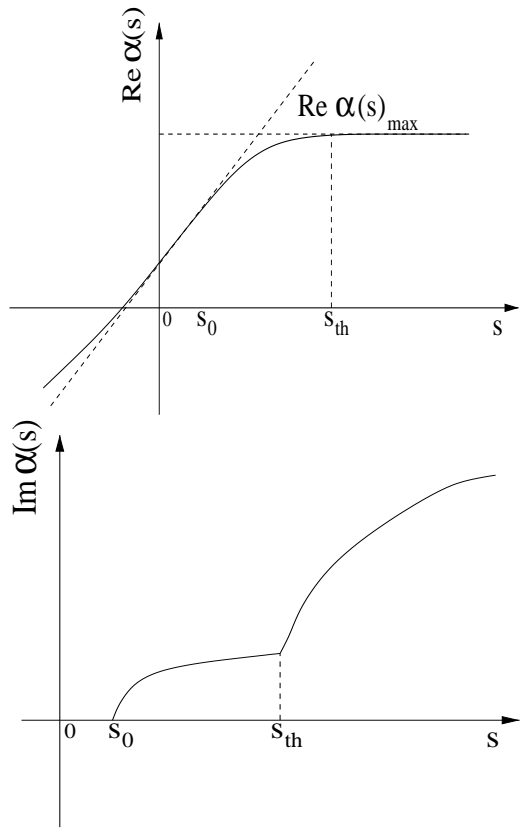}
\caption{Typical
behavior of the real and imaginary parts of
an analytic model of the Regge trajectory \cite{D}. The (nearly) smooth
behaviour results from the smearing of several additive 
thresholds in Eq.~(\ref{m19}).
} 
\label{tr}
\end{figure}

Explicit models of Regge trajectories realizing the above
requirements were studied in a number of papers (see Ref.~\cite{FJMPP}
and references therein). The main feature of these trajectories is
that the number of resonances is finite (due to an upper limit on their real
part, determined by fits to the data), as illustrated in Fig. \ref{tr}.
 A particularly simple model is based on a sum of square roots 
\beq
\alpha(s)=\alpha_0+\sum_i\gamma_i \sqrt{s_i-s}, 
\eeq{m19}
where the lightest threshold gives rise to the imaginary part
while the heaviest one promotes the nearly linear rise of the real
part at small and intermediate $s$. To simplify the calculations,
in a limited range of $s$, the heaviest threshold can be
approximated by a linear term, hence the expression (\ref{m19}) for 
the trajectory can be
written as
\beq
\alpha(s)=\alpha_0+\alpha_1 s+\alpha_2(\sqrt {s_0}-\sqrt{s_0-s}),
\eeq{m18}
where $s_0$ is the lightest threshold, $s_0=(m_{\pi}+m_p)^2=1.14$ GeV$^2$ in our
case. Beyond the threshold, the real part of this trajectory is
${\cal R}e\ \alpha(s)=\alpha_0+\alpha_2\sqrt {s_0}+\alpha_1 s$, 
while its imaginary part is 
${\cal I}m\ \alpha(s)=\alpha_2\sqrt{s-s_0}.$

Large-angle scaling behavior of the hadronic amplitudes
constrains \cite{JMP1} the asymptotic (far away from the resonance
region) behavior of the trajectories by a logarithm .

In $\gamma^* p$ scattering only $N$ (isospin 1/2)
and $\Delta$ (isospin 3/2) resonances contribute in the $s$ channel.
 The latest
Review of Particle Physics \cite{PDG} quotes about 14 $N$'s whose
``existence is certain or very likely certain'' plus 8 $N'$s whose
``evidence is fair or poor'') and 10 ``certain or likely certain''
(plus 12 ``fair or poor'') $\Delta$'s. Most, if not all of the
above two dozen (almost) certain resonances - among them the
``prominent'' ones mentioned in the Introduction - contribute, with
different weights, to the the $\gamma^* N$ total cross section or to
the nucleon SF. It is clear that a systematic account for all
these resonances (plus those to be confirmed) is not an easy task.
A much more economic way is to introduce the whole sequence of
recurrences. As well as generalizing the concept of a resonance (a
Regge trajectory realizes the analytic continuation of the
discrete resonance spin and is an indispensable ingredient of dual
models!), the trajectory may also be used to classify the
resonances by eliminating some candidates and predicting others.
The above resonances lie on several positive- and negative-parity $N$
and $\Delta$ trajectories.

The classification of meson and baryon resonances on the basis of
Regge recurrences of SU(3) multiplets was extensively studied in
the late sixties (see e.g. Ref.~\cite{Collins}). A nontrivial model for
meson Regge trajectories with the limited real part was suggested
in a recent paper \cite{FJMPP}. Anticipating analogous new
results for baryon trajectories, below we shall use a simple model,
based on the ideas
introduced above and fitted to the three ``prominent'' resonances,
seen in $ep$ scattering.

\section {The model}\label{s4}

To illustrate these ideas, we start with a simplified model, in
which we disregard the helicity structure of the scattering
amplitude and relevant selection rules, concentrating on the role
of the Regge trajectories, analyticity and duality. Their role was
treated in a number of papers \cite{CM} (recently in Ref.~\cite{DS}).

In the dual-Regge approach \cite{JMP,JMP-1,JMSz} the Compton
scattering can be viewed as an off-mass shell continuation of a
hadronic reaction, dominated in the resonance region by
non-strange ($N$ and $\Delta$) baryon trajectories. The scattering
amplitude follows from the pole decomposition of a dual amplitude
\cite{JMP}
\beq
A(s,Q^2)\Biggl|_{t=0}= norm\sum_{i=N_1^*,N_2^*,\Delta,E}A_i
\sum_{n=n_i^{min}}^{n_i^{max}}{f_{i}(Q^2)^{2\left(n-n_i^{min}+1\right)}
\over{n-\alpha_{i}(s)}}~,
\eeq{m17}
where $i$ runs over all the trajectories allowed by quantum number exchange, 
$norm$ and $A_i$'s are constants, $f_i(Q^2)$'s are the form factors. 
These form factors generalize 
the concept of inelastic (transition) form factors to the case 
of continuous spin,
represented by the direct-channel trajectories. The $n_i^{min}$ refers
to the spin of the first resonance on the corresponding trajectory $i$
(it is convenient to shift the trajectories by $1/2$,
therefore we use  $\alpha_i=\alpha^{phys}_i-1/2$, which due
to the semi-integer values of the baryon spin leaves $n$ in 
Eq.~(\ref{m17}) integer).
The sum
over $n$ goes with step $2$ (in order to conserve parity).

It follows from Eq.~(\ref{m17}) that
\beq
{\cal I}m\ A(s,Q^2)=norm\sum_{i=N_1^*,N_2^*,\Delta,E}A_i
\sum_{n=n_i^{min}}^{n_i^{max}}{[f_i(Q^2)]^{2\left(n-n_i^{min}+1\right)}
{\cal I}m\ \alpha_i(s)
\over {(n-{\cal R}e\ \alpha_i(s))^2+\left({\cal I}m\ \alpha_i(s)\right)^2}}\ .
\eeq{sumsum}

Eqs.~(\ref{m17}, \ref{sumsum}) have a factorized form, a 
product of vertices (two identical form factors) times the propagator.
Each term in Eq.~(\ref{sumsum}) resembles that of a 
Breit-Wigner formula with a
resonance mass $s_n=(n-\alpha_0-\alpha_2\sqrt {s_0})/\alpha_1$ 
(${\cal R}e\ \alpha(s_n)=n$) and an
energy-dependent width given by ${\cal I}m\ \alpha(s_n)/\alpha_1$ of the
trajectory. The factor
$(n-\alpha(s))^{-1}$ is typical of the dual-Regge approach \cite{JMP}.

The first three terms in (\ref{sumsum})
are the non-singlet, or Reggeon contributions
with the $N^*$ and $\Delta$  trajectories in the  $s$-channel, dual to the
exchange  of an effective bosonic trajectory (essentially, $f$) in the
$t$-channel, and the fourth term is the contribution from the smooth
background, modeled by a non-resonance pole term with an exotic trajectory
$\alpha_E(s)$, dual to the pomeron (see Ref.~\cite{JMP}). As argued in 
Ref.~\cite{JMP}, only a limited number, ${\cal N}$, 
of resonances appear on the trajectories,
for which reason we tentatively set ${\cal N}=3$, i.e. one resonance on
each trajectories ($N^*_1,\ N^*_2,\ \Delta$). 
We tried also with higher values of ${\cal N}$, up to ${\cal N}=10$, but our analyses  
shows that ${\cal N}=3$ is a
reasonable approximation --
even if additional peaks appear, they are suppressed with respect 
to the dominant one (first on each trajectory), because of the $Q^2-$behaviour of the form 
factors. 
Thus, the limited (small) number of resonances 
contributing to the cross section results not only from the termination of
resonances on a trajectory but even more due to the strong suppression
coming from the numerator (increasing powers
of the form factors).

Since, by definition, the smooth background does not show any
resonance, here we keep only one term in the sum (for more details
see \cite{JMP,JMP-1,JMSz} and references therein).

By inserting in Eq.~(\ref{sumsum}) the relevant baryonic trajectories, we
include all possible electromagnetic transitions from a nucleon to
its excitations. There are, however, selection rules, enhancing some
and suppressing others. This effect will be taken into account
by our phenomenological fits to the data.

As discussed above the nonlinear behaviour of the
trajectories, especially the boundedness of their real part, is a
crucial feature of our dual model.  For practical reasons we have
replaced the formal condition ${\cal R}e\ \alpha(s)$ $< const$ by a finite
sum in Eq.~(\ref{sumsum}), introducing a linear term in the baryon 
trajectory to
approximate the contribution from heavy thresholds (see Eq.~(\ref{m18})).

We have fitted the parameters of the baryon trajectories, 
given by Eq.~(\ref{m18}), 
such as to reproduce the experimental masses and
widths of the following resonances:\\ 

\begin{tabular}{c|c|c|c}
Resonance & $I\ \left(J^P\right)$ & $M,$ GeV & $\Gamma,$ GeV\\
\hline
$\Delta(1236)$ & $\frac{3}{2}\left(\frac{3}{2}^+\right)$ & $1.230$ & $0.115$\\
\hline
$N^*(1520)$  & $\frac{1}{2}\left(\frac{3}{2}^-\right)$ & $ 1.515$ & $0.11$\\
\hline
$N^*(1680)$  & $\frac{1}{2}\left(\frac{5}{2}^+\right)$ &  $1.690$ & $0.12$\\
\end{tabular}\\

We end up with the following trajectories:
\beq
\alpha_{N_1^*}(s)=-0.8377+0.95s+0.1473(\sqrt{s_0}-\sqrt{s_0-s})\ ,
\eeq{Nstar1} \beq
\alpha_{N_2^*}(s)=-0.37+0.95s+0.1471(\sqrt{s_0}-\sqrt{s_0-s})\ ,
\eeq{Nstar2} \beq
\alpha_{\Delta}(s)=0.0038+0.85s+0.1969(\sqrt{s_0}-\sqrt{s_0-s})\ .
\eeq{Delta}

We take only one resonance on each trajectory, so
the sum over $n$ in Eq.~(\ref{sumsum}) reduces  to one term, i.e. 
$n_i^{max}=n_i^{min}$ for all three baryon trajectories as well 
as for the exotic one.

We take the exotic trajectory in the form
\begin{equation}
\alpha_E(s)=\alpha_E(0)+\alpha_{1E}(\sqrt{s_E}-\sqrt{s_E-s}),
\end{equation}
where the intercept $\alpha_E(0)$, $\alpha_{1E}$ and the effective 
exotic threshold $s_E$
are free parameters. 
As a first approximation we can assume the following expression for
the exotic trajectory \cite{JMP}:
\beq
\alpha_E(s)=0.5+0.12(\sqrt{s_E}-\sqrt{s_E-s})\ ,
\eeq{exotic}
where $s_E=1.145^2$ GeV$^2$. 
As we said
above, we take only one term from exotic trajectory. 
$n_E^{min}$ is the first integer larger then 
$Max({\cal R}e\ \alpha_E)$ (to make sure there are no
resonances on the exotic trajectory); in our case $n_E^{min}=1$.

In dual models, the numerator in Eq.~(\ref{sumsum}) contains powers of
the intercepts of the $t$-channel trajectories, $\alpha_t^n(0)$
and/or powers of a parameter $g$ that in Ref.~\cite{JMP} was
assumed to be $Q^2$-dependent by matching the Regge behaviour and the 
Bjorken scaling. As it was mentioned before, here we invert
the problem and, in accordance with the factorization properties
of the amplitude, we multiply the resonance propagators by the
product of the inelastic form factors (to be later identified with the
parameters of the dual model \cite{JMP}). 

To start with we  use the
simplest, dipole model for the form factors, disregarding 
the spin structure of the amplitude and the difference between electric
and magnetic form factors:
\beq
f_i(Q^2)={1\over (1+{Q^2\over Q^2_{0,i}})^{2}}\ .
\eeq{ff}
where $Q_{0,i}^2$ are scaling parameters.

From Eq.~(\ref{sumsum}) one can immediately guess that the
first resonance on each trajectory, $\Delta,\ N_1^*,\ N_2^*$ or
$E$, will become dominant over the subsequent ones with increasing
$Q^2$ due to the power behavior of the form factors. The relative growth 
of these three terms will depend on the scaling factor $Q^2_{0,i}$.
Therefore we  choose $Q^2_{0,E}>Q^2_{0,{N_2^*}}>Q^2_{0,{N_1^*}}>Q^2_{0,\Delta}$ in
order to satisfy the experimentally observed behaviour of these
terms, for example, the rise of the background contribution with
respect to the resonance one with increasing $Q^2$; the relative growth
of the $N_1^*$ and $N_2^*$ peaks with respect to the $\Delta$ peak.

\section {Fits to the SLAC and JLab data}\label{s5}

In this Section we present a numerical analysis of our model based on the 
experimental data from SLAC \cite{Stoler} and JLAB \cite{SLAC}\footnote{We are grateful to 
M.I. Niculescu for making her data compilation available to us.}. 

Before displaying the results of our fits, it is instructive to illustrate 
the ideas behind our study, in order to make the following results 
physically more transparent.

We are dealing with a set of 704 experimental points for
7 different values of $Q^2$: 0.45, 0.85, 1.4, 1.7, 
2.4, 3.0, 3.3 GeV$^2$. A first rough fit when the whole data-set
is taken into account produces $\chi^2_{d.o.f.}\propto 10^4$.
This has to do with the fact that the set of experimental data
is not homogeneous, i.e. points at low $s$
(high $x$) are given with  very small experimental errors,
thus ``weighting'' the fitting procedure not uniformly. 
This forced us to make a preselection
in the initial data-set by removing points with 
$\chi^2_{red}\le 1000$. 
Below we will be considering the leftover 634 data 
points (90\% of the original data-set), although all the experimental 
points are presented in the Figures.
 
\begin{figure}[htb]
\vspace*{-1.0cm}
        \insertplotlll{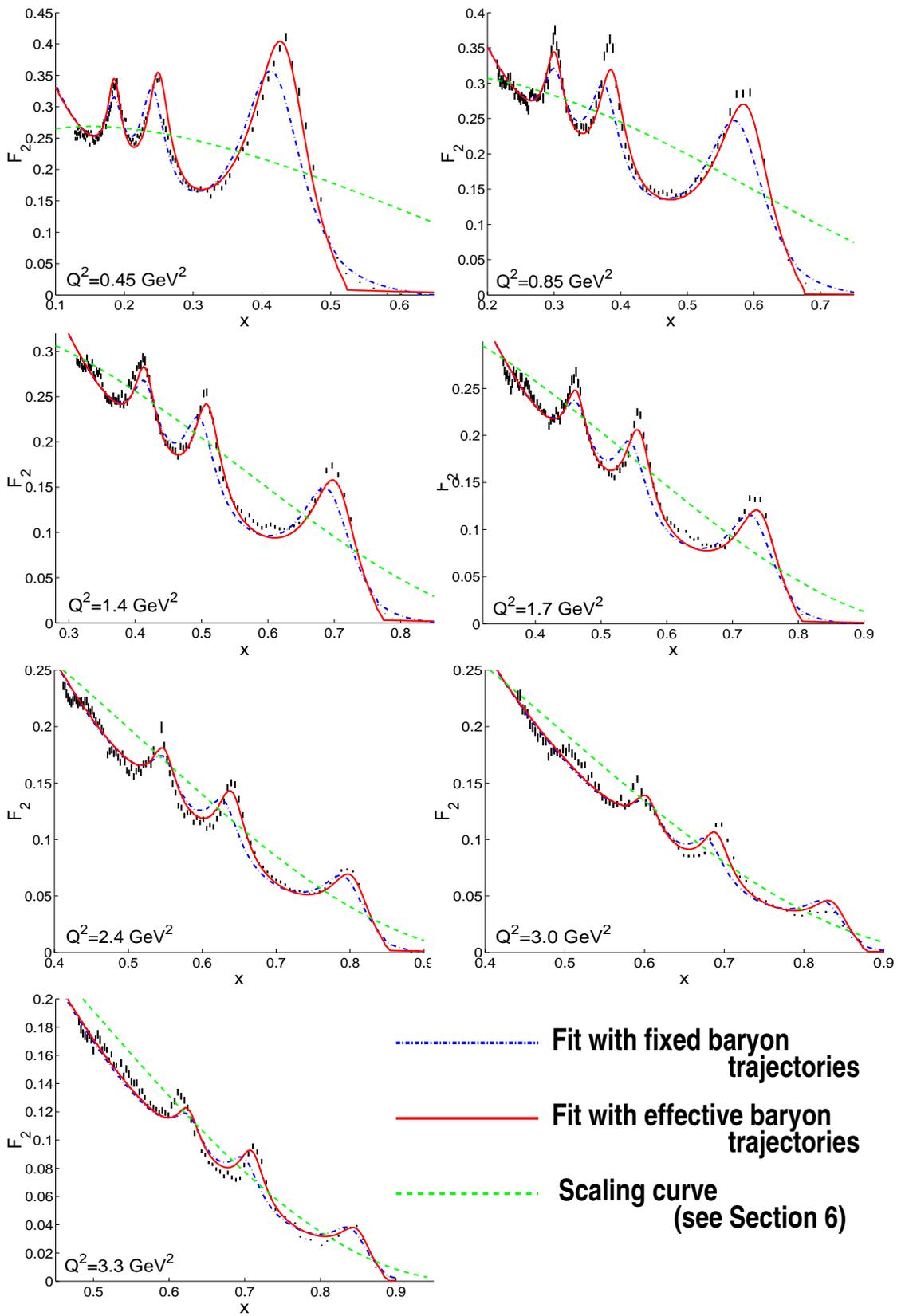}
\vspace*{-0.5cm}
\caption{$F_2$ as a function of $x$ for $Q^2 = 0.45-3.3$ GeV$^2$. } 
\label{fit1} 
\end{figure}

As we have seen in Section \ref{s4}, our model consists of the sum of the 
contributions from three resonances ($N^{*}_{1}$, $N^{*}_{2}$, 
$\Delta$), which give the dominant contribution to 
the structure functions, plus an exotic background, that we treat 
effectively. The first approach to the 
fitting procedure consists in 
fixing parameters of the trajectories for the resonances, in order to 
reproduce the correct masses and widths, leaving the four scaling 
constants $Q^2_i$, four factors $A_i$ and the parameters of the exotic 
trajectories to be fitted to the data.
The results are shown in Table \ref{p1} (first coloumn) 
and the plots of the SF against $x$ are 
presented in Fig. \ref{fit1} (dashed-dotted lines).

Although the agreement with the experimental data seems not to be good 
($\chi^2_{red}=28.29$), 
there is a number of features which we could introduce in order 
to improve the model.

A first important aspect is to account for the large number 
of resonances (about 20) present in the energy range under investigation, 
which overlap, as noted in Section \ref{s3}. 
For this reason we consider the dominant resonances 
($N^*_1$, $N^*_2$ and $\Delta$) as ``effective'' contributions to the SF.
In other words we require that they mimic the contribution of 
the dominant resonances plus the large number of subleading 
contributions, which, together, fully describe the real physical system.

A way of doing this is to consider corrections to the model as explained 
in the following. 
If we denote the structure function as 
$F(\alpha,x,Q^2)$, where $\alpha$ represents generically the 
trajectories' parameters, 
the contribution from the subleading resonances can be introduced as 
corrections to the $\alpha$'s. Thus, the ``effective'' SF becomes a function 
of some corrected trajectories
$F(\alpha + \delta \alpha,x,Q^2)$, where $\alpha$, represents the physical 
value and $\delta \alpha$ the correction which accounts for the 
subleading resonances. However, due to the fact 
that the physical resonances give the dominant contribution, we expect 
this departure not to be large. 

In the light of these considerations, we have refitted the data, 
allowing the baryon trajectories parameters to vary. The resulting 
parameters of such a fit 
are reported  in Table \ref{p1} (second coloumn). 
It is worth noting that although the range of variation was not restricted,
the new parameters of the trajectories stay close to their physical values,
showing stability of the fit and thus reinforcing our previous considerations. 
From the relevant plots, shown in Fig. \ref{fit1} with full lines,
one can see that 
the improvement is significant, although agreement is still far from being 
perfect ($\chi^2_{d.o.f.}=11.6$). 

Another important ingredient to be introduced in the model is spin. This 
changes the form factors in a non-trivial way, thus complicating the 
$Q^2-$dependence of the SF's (see Ref.~\cite{DS} for a recent treatment of 
the problem). 
These corrections have not yet been included in our study and might be responsible for 
relatively poor agreement with data. We hope to address this problem 
in a forthcoming work. 

One may also ask the question of how good the dipole expression for the form factors, 
Eq.~(\ref{ff}) does work. To answer this question we performed the fit 
letting the powers of the
$1/(1+Q^2/Q^2_{0,i})$ in Eq.~(\ref{ff}) free to vary. The results show 
that second power is a 
good approximation - powers change only by about $5\% $. As we shall see
in the next Section, the dipole approximation deteriorates towards large
values of $Q^2$. This phenomenon may be partly due to the spin
effects, ignored in the present model.

\section {Parton-hadron duality: a numerical tests}
\label{s7}

In order to quantify the validity of the quark-hadron duality, it is 
customary (see e.g. Ref.~\cite{Dissertation}) to compare the $Q^2 
-$behaviour of the following 
quantities:
\beq
 I_{Bj}(Q^2)=\int_{s_{th}}^{s_{max}} ds\ F_2^{Bj} \ , 
\eeq{n21} 
\beq
I_{Model}(Q^2) = \int_{s_{th}}^{s_{max}} ds\ F_2^{Model}\ , 
\eeq{n22}
where the lower integration limit is fixed to $s_{th}=s_0$ and 
the upper integration limit, $s_{max}$, is varied within the range $3-25$ 
GeV$^2$.

Here $F_2^{Model}$ is our model, given by Eqs.~(\ref{m23}), 
(\ref{m22}), (\ref{sumsum}) and $F_2^{Bj}$ is a ``scaling curve'', i.e. a phenomenological 
parameterizations of the SF exhibiting Bjorken scaling and fitting the data. 
We have chosen the parameterizations studied in Ref. \cite{BCDMS}, which 
we report for the convenience of the reader,
\begin{equation}
\label{eq5} F_{2}^{Bj}(x,Q^{2})=F_{S}(x,Q^{2})+F_{NS}(x,Q^{2})\ ,
\end{equation}
where
\begin{equation}
\label{eq6} F_{NS}(x,Q^{2})=D\cdot x^{1-\alpha_{R}}\cdot
(1-x)^{n(Q^{2})}\cdot
\left(\frac{Q^{2}}{Q^{2}+b}\right)^{\alpha_{R}}\ ,
\end{equation}
\begin{equation}
\label{eq7} n(Q^{2})=\frac{3}{2} \cdot
\left(1+\frac{Q^{2}}{Q^{2}+c}\right)\ .
\end{equation}
The singlet component of the SF, corresponding to a multipole \linebreak
(single+double+triple) Pomeron is a sum of logarithms:
\begin{equation}
\begin{array}{cc}
\label{eq8} F_{S}(x,Q^{2})=Q^{2}
\left[\right. A\left(\frac{a}{a+Q^{2}}\right)^{\alpha}+B\left(\frac{a}
{a+Q^{2}}\right)^{\beta} \log{\left(\frac{Q^{2}}{x}\right)}+\\ 
C\left(\frac{a}{a+Q^{2}}\right)^{\gamma}
\log^2{\left(\frac{Q^{2}}{x}\right)}\left.\right] (1-x)^{n(Q^{2})+4}\ .
\end{array}
\end{equation}
The values of the parameters, $A,\ B,\ C,\ D,$ $a,\ b,\ c,$ $\alpha,\ \beta,\ \gamma$ and $\alpha_R$ 
can be found in \cite{BCDMS}.
The corresponding curves are presented by dashed lines in Fig. (\ref{fit1}).

\begin{figure}[htb]
        \insertplot{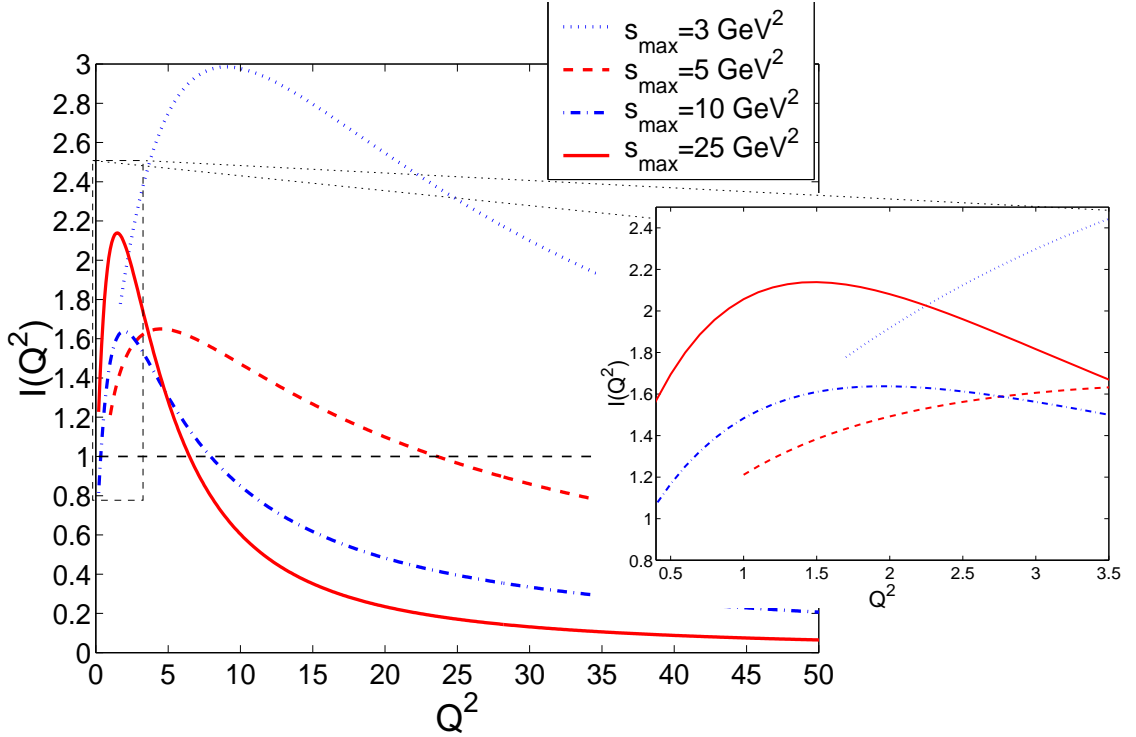}
\caption{Parton-hadron duality test for different $s_{max}$.} 
\label{smax}
\end{figure}

The quantity of interest is the so called duality ratio given by 
\beq
{\cal I} (Q^2) = {I_{model}(Q^2) \over I_{Bj}(Q^2)}\ .
\eeq{nrat}
Its deviation from unity is indicative of any violation of
parton-hadron duality.
We compute numerically this quantity, using values of the second set of
parameters 
(Table (\ref{p1}), second coloumn) and show the result in Fig. \ref{smax}. 

One can see from Fig. \ref{smax} that for a short interval of integration
our model strongly overshoots the ``scaling curve'', since in this region we have 
strongly peaked resonances.
For $s_{max}$ larger than 5 GeV$^2$ this effect saturates.
As we tend to the large $Q^2$ region our model starts to underestimate 
the ``scaling curve''. It was 
stressed in Ref.~\cite{BCDMS} that the 
scaling curve itself starts to overestimate the data for $Q^2>15-20$ GeV$^2$. 
The strong decrease of ${\cal I}(Q^2)$ for high $Q^2$ is partly due to this
fact. 
The subplot shows that in the $Q^2$ interval of interest the ${\cal I}(Q^2)$ 
deviates from 1 by 20-40\%. 

\begin{figure}[htb]
        \insertplotll{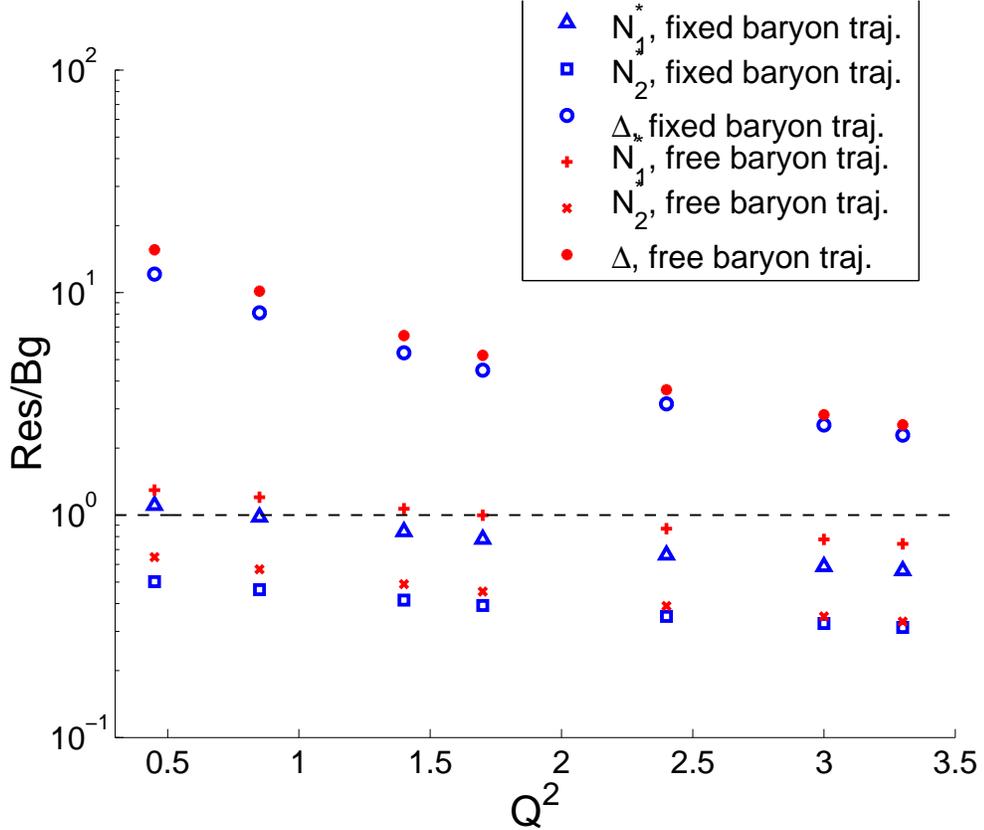}
\caption{The ratio of the resonance to
 background components of the SF at the resonance peaks. 
 See text for more details.} 
\label{rat}
\end{figure}

We have calculated also the $Q^2$-dependent ratio of the resonance to
background components of the  SF at three fixed values x, namely at
three physical 
resonance peaks, $s_{N_1^*}$, $s_{N_2^*}$, $s_{\Delta}$, for the fit 
with fixed physical baryon trajectories and at effective resonance peaks,
$s_{N_1^*}^{eff}=1.494^2$ GeV$^2$, $s_{N_2^*}^{eff}=1.690^2$ 
GeV$^2$, $s_{\Delta}^{eff}=1.22^2$ GeV$^2$, for the fit with free baryon 
trajectories\footnote{Notice that the position of the $N_2^*$ resonance 
remains the same 
for the effective trajectory. Others also do not change much.}. 
On this plot  
the ``background'' for the selected resonance consists of three parts, i.e. 
contribution from the 
exotic trajectory (usual background term) and contributions from 
the two other resonances.  
The results are shown in Fig. \ref{rat}. One may see that for $N_2^*$ 
and for $N_1^*$ for $Q^2>1.5$ GeV$^2$ the ``background'' contributes more than 
the resonant term itself.

\section {Parton-hadron duality: an analytical tests}\label{s8}

Consider now the behaviour of $F_2(x,Q^2)$ at large $x$ when  $s$
is kept in the resonance region.
Let us remind the reader that $x, Q^2$ and $s$ are related by
Eq.~(\ref{m21}) with $m=m_p$.
Thus, each term in the rhs
of Eq.~(\ref{sumsum}), using Eqs.~(\ref{m23}, \ref{m22}) looks like
$$
F_2(x,Q^2)_{i,n}={Q^2(1-x)\over{4\pi \alpha (1+{4m_p^2 x^2\over Q^2})}}
\frac{norm\ A_i}{(1+{Q^2\over Q^2_{0,i}})^{4(n-n_i^{min}+1)}}
$$
\beq 
\cdot {{\cal I}m\ \alpha_i(s) \over {(n-{\cal R}e\ \alpha_i(s))^2+({\cal I}m\
\alpha_i(s))^2}}\ . 
\eeq{t1} 
In our case, since we consider only one resonance on each trajectory,
we have 
\beq
F_2(x,Q^2)_{i,n}\propto \frac{1}{(1+{Q^2\over Q^2_{0,i}})^4}~.
\eeq{t1pr}
In the limit of $x$ going to $1$ and $s$
in the resonance region ($1-4$ GeV$^2$),
$Q^2=x(s-m_p^2)/(1-x)$ is much larger than $s$ and $Q^2_{0,i}$, which
are of the same order. Therefore 
\beq
\frac{1}{(1+{Q^2\over Q^2_{0,i}})^{4}}\approx
\frac{(1-x)^{4}}{\left(x(s-m_p^2)\over
Q^2_{0,i}\right)^{4}}
\left(1-4{Q^2_{0,i}\over Q^2}+O\left({1\over Q^4}\right)\right)\ . 
\eeq{formlim}
Using $x=Q^2/(Q^2+s-m_p^2)=1-(s-m_p^2)/Q^2$ we can go one step
further: 
\beq \frac{1}{(1+{Q^2\over Q^2_{0,i}})^{4}}\approx
\frac{(1-x)^{4}}{\left((s-m_p^2)\over
Q^2_{0,i}\right)^{4}}
\left(1-4\frac{Q^2_{0,i}+s-m_p^2}{Q^2}+O\left({1\over Q^4}\right)\right)\ .
\eeq{formlim2} 
Thus
$$
F_2(x,Q^2)\approx norm\sum_{i=N_1^*,N_2^*,\Delta,E}A_i
\sum_{n=n_i^{min}}^{n_i^{max}}(1-x)^{4}
$$
\beq
\cdot M_{i,n}(x,Q^2) \left(1-4\frac{Q^2_{0,i}+s-m_p^2}{Q^2}+O\left({1\over Q^4}\right)\right), 
\eeq{f2lim} 
where \beq M_{i,n}(x,Q^2)= {(s-m_p^2)x \over
4\pi \alpha (1+{4m_p^2x^2\over Q^2})} \left(Q^2_{0,i} \over
{s-m_p^2}\right)^{4} {{\cal I}m\ \alpha_i(s) \over {(n-{\cal R}e\
\alpha_i(s))^2+({\cal I}m\ \alpha_i(s))^2}}~. \eeq{Ms} 
In our range of interest
$M_{i,n}$ is a slowly varying function of both $x$ and $Q^2$. For each
$(i,n)$ the term proportional to 
$(1-x)^{4(n-n_i^{max}+1)}$ shows the main tendency of
$F_2(x,Q^2)_{i,n}$, while $M_{i,n}$ is responsible for the
``fine structure'' - resonances at large $x$. Of course, for each
trajectory $i$ the main contribution comes from the first
resonance - $(1-x)^{4}$.

	At this point it might be interesting to see 
the effect of spin corrections. As 
it has been shown in Ref.~\cite{DS}, if one explicitly takes  	
into account the spin structure of the $F_2$, the main contribution 
from each resonance in the limit $x\rightarrow 1$ 
($Q^2\rightarrow\infty$) is proportional 
to $(1-x)^{3}$. Thus our model, neglected spin effects, strongly 
underestimate the
physical SF. This might be another reason why the duality ration, 
Eq.~(\ref{nrat}), is so small in the high $Q^2$ region.

\section {Conclusions}\label{s9}

The idea of the present paper is that deep inelastic scattering
can be described by a sum of direct channel resonances lying on
Regge trajectories. The form of these trajectories is crucial for
the dynamics. It is constrained by analyticity, unitarity and by
the experimental data. 
The use of baryon trajectories instead of individual resonances
not only makes the model economic (several resonances are replaced
by one trajectory) but also helps in classifying the resonances,
by including the ``right'' ones and eliminating those nonexistent.

   Compton scattering of nuclei is much similar to $\pi N$
scattering. The difference coming mainly from the photon spin
imposes selection rules on inelastic transitions. These
selection rules are approximate, and can be introduced either in
the construction of the scattering amplitude (cross section,
SF), or just by fitting the model to the data.

To fix the ideas and make a rough fit to the data, we constructed a
simplified model with just $3$ baryon trajectories, in which 
heavy thresholds have been replaced for
simplicity by a linear term, and
with the lowest-lying resonances. In fact, apart from the
``prominent'' three resonances many more should be included
by means of relevant baryon trajectories. To this end an
independent study of baryon trajectories and updated fits to
dozens of existing resonances should be done. We intend to
continue working in this direction.

The dynamics of DIS can be described either by the contribution of
direct-channel or cross-channel Regge trajectories.

\begin{figure}[htb]
        \insertplotsshort{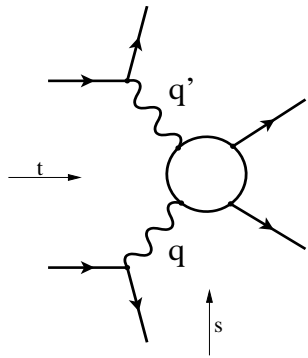}
\caption{Typical six-point function ansatz.} 
\label{6p}
\end{figure}

An important issue of our paper is the integral representation
realizing parton-hadron, or Bloom-Gilman duality. This can be seen
from the equivalence of its pole decomposition (resonances)
and smooth large-$Q^2$  scaling behavior. As a result, a $Q^2-$
dependent photon Regge trajectory emerges. In the unphysical region of
positive $q^2$ (negative $Q^2$), outside the one we consider,
resonances are expected to appear on this trajectory. As it is known,
these are the vector mesons  $\rho, \omega, \phi, J/\Psi$ etc. The
non-appearance of higher recurrences, with spin larger than $1$, on 
the photon
trajectory is a problem for linear photon trajectories.
``Analytic'' models of Regge  trajectories used in our paper can
resolve this problem: the real part of the photon trajectory is
limited by ${\cal R}e\ \alpha_\gamma(Q^2)<2$ and  higher than
spin one states are not expected there.

The dual model  
presented here is a part of a more
general 6-point dual amplitude (see Fig. \ref{6p}), where two pairs of
external particles (electrons) are at a pole, corresponding to the
photon trajectory $\alpha_\gamma(Q^2)$. Such a dual model, but without
the important limitation ${\cal R}e\ \alpha_\gamma(Q^2)<2$ and for $t=0$, 
was
considered in Ref.~\cite{Sch}. Relaxing the condition $t=0$ one
can apply the 6-point dual amplitude with the above trajectories
to DIS in all possible kinematical regions, including off-diagonal
or skew-symmetric (i.e. where $t\neq 0$) DIS. We intend to report on
relevant results in a forthcoming publication.

\vskip 0.2cm

{\bf Acknowledgments}

We thank F. Paccanoni, E. Predazzi and B.V. Struminsky for
fruitful discussions and I. Niculescu for a very useful and
stimulating correspondence. L.J. and V.M., acknowledge the 
support from INTAS, grant 00-00366. A.F. and V.M. wish to thank the 
University of Calabria, where this work was completed, for 
the warm hospitality.

\begin{table}[htb]
\caption{Parameters of the fit. In the first coloumn we show the result of 
the fit when the parameters of the barionic trajectories are fixed. The 
second coloumn contains the result of the fit when the parameters of the 
trajectories are  varied. $^\dagger$ - parameters of the physical 
baryon trajectories 
from Eqs.~(\ref{Nstar1}-\ref{Delta}).
$^*$ - the coefficient $norm$ is chosen in such a way as 
to keep $A_{N_1^*}=1$ in order to see 
the interplay between different resonances. 
} 
\vspace{0.5cm}
\begin{tabular}{|c|c|c|c|}
\hline
  &$\alpha_{0}$                   & -0.8377 (fixed)$^\dagger$  & -0.8070 \\
  &$\alpha_{1}$                   &  0.95  (fixed)$^\dagger$  &  0.9632 \\
$N_1^*$           & $\alpha_{2}$  &   0.1473 (fixed)$^\dagger$ & 0.1387  \\
  &$A_{N_1^*}$                    &  1 (fixed)$^*$&  1 (fixed)$^*$ \\
  &$Q^2_{N_1^*}$, GeV$^2 $                  &  2.4617& 2.6066 \\
\hline
\hline
  &$\alpha_{0}$                   &  -0.37(fixed)$^\dagger$  & -0.3640\\
  &$\alpha_{1}$                   &   0.95  (fixed)$^\dagger$ &  0.9531 \\
$N_2^*$           & $\alpha_{2}$  &   0.1471 (fixed)$^\dagger$ & 0.1239\\
 & $A_{N_2^*}$                   &  0.5399& 0.6086\\
  & $Q^2_{N_2^*}$, GeV$^2 $                 &   2.9727& 2.6614 \\
\hline
\hline
  &$\alpha_{0}$                   &  0.0038 (fixed)$^\dagger$  & -0.0065 \\
  &$\alpha_{1}$                   &  0.85    (fixed)$^\dagger$ & 0.8355\\
$\Delta$          & $\alpha_{2}$  &  0.1969  (fixed)$^\dagger$ &  0.2320\\
  & $A_{\Delta}$                  &   4.2225&  4.7279\\
  & $Q^2_{\Delta}$, GeV$^2 $                &   1.5722 & 1.4828 \\
\hline
\hline
  & $s_{0}$, GeV$^2 $                    &  1.14 (fixed)$^\dagger$   & 1.2871 \\
\hline
\hline
  &$\alpha_{0}$                   &  0.5645  &  0.5484\\
${E}$        & $\alpha_{2}$  &  0.1126    & 0.1373 \\
  & $s_{E}$, GeV$^2 $                    &  1.3086  & 1.3139 \\
  & $A_{exot}$                    &  19.2694  &  14.7267\\
  & $Q^2_{exot}$, GeV$^2 $                  &  4.5259  &  4.6041\\
\hline
\hline
& $norm$ & 0.021 & 0.0207\\
\hline
\hline
\hline
$\chi^2_{d.o.f.}$&                   &  28.29    & 11.60 \\
\hline
\hline
\end{tabular}
\label{p1}
\end{table}

\vfill \eject

\end{document}